\documentclass[
  twoside,
  twocolumn,
  aps,
  prb,
  preprintnumbers,
  floatfix,
  citeautoscript,
  10pt,
]{revtex4-1}

\usepackage{amsmath}
\usepackage{amssymb}
\usepackage{color}
\usepackage{xcolor}
\usepackage{graphicx}
\usepackage{tabularx}
\usepackage{dcolumn}
\usepackage{times}
\usepackage{units}
\usepackage{float}

\definecolor{chalmersblue}{rgb}{0,0,.6}
\usepackage[bookmarks=true,
            colorlinks=true,
            linkcolor=chalmersblue,
            urlcolor=chalmersblue,
            filecolor=chalmersblue,
            citecolor=chalmersblue,
            linktoc=all]{hyperref}

\renewcommand{\S}{\text{S}}
\providecommand{\widebar}[1]{\bar{#1}}
\newcommand{\Se}{\text{Se}}
\newcommand{\Te}{\text{Te}}

\newcommand{\sect}[1]{Sect.~\ref{#1}}
\newcommand{\fig}[1]{Fig.~\ref{#1}}
\newcommand{\eq}[1]{Eq.~(\ref{#1})}
\newcommand{\tab}[1]{Table~\ref{#1}}

\renewcommand{\vec}[1]{\ensuremath\boldsymbol{#1}}
\renewcommand{\epsilon}{\varepsilon}

\newcommand{\myscale}{0.65}

\begin{document}

\title{
  Thermal expansion and transport in van der Waals solids from first-principles calculations
}

\author{Daniel O. Lindroth}
\email{daniel.lindroth@chalmers.se}
\author{Paul Erhart}
\email{erhart@chalmers.se}
\affiliation{
  Chalmers University of Technology,
  Department of Physics,
  Gothenburg, Sweden
}

\begin{abstract}
The lattice thermal expansion and conductivity in bulk Mo and W-based transition metal dichalcogenides are investigated by means of density functional and Boltzmann transport theory calculations. To this end, a recent van der Waals density functional (vdW-DF-CX) is employed, which is shown to yield excellent agreement with reference data for the structural parameters. The calculated in-plane thermal conductivity compares well with experimental room temperature values, when phonon-phonon and isotopic scattering are included. To explain the behavior over the entire available temperature range one must, however, include additional (temperature independent) scattering mechanisms that limit the mean free path. Generally, the primary heat carrying modes have mean free paths of $1\,\mu\text{m}$ or more, which makes these materials very susceptible to structural defects. The conductivity of Mo and W-based TMDs is primarily determined by the chalcogenide species and increases in the order Te-Se-S. While for the tellurides and selenides the transition metal element has a negligible effect, the conductivity of WS$_2$ is notably higher than for MoS$_2$, which can be traced to the much larger phonon band gap of the former. Overall the present study provides a consistent set of thermal conductivities that reveal chemical trends and constitute the basis for future investigations of van der Waals solids.
\end{abstract}


\maketitle

\section{Introduction}

In the advent of increasingly elaborate synthesis techniques \cite{GeiGri13, GonLinWan14} highly engineered van der Waals (vdW) solids are emerging as promising candidates for a manifold of applications including electronic components \cite{RadRadBri11}, optoelectronics \cite{WanKalKou12, HonKimShi14, MasSchVia16}, thermoelectrics \cite{GuoYanTao13}, and spintronics \cite{Han16}. Since thermal transport plays a key role in many of these situations, it is important to develop a detailed understanding of the thermal conductivity in vdW solids.

\begin{figure}[H]
  \centering
\includegraphics[width=0.9\linewidth]{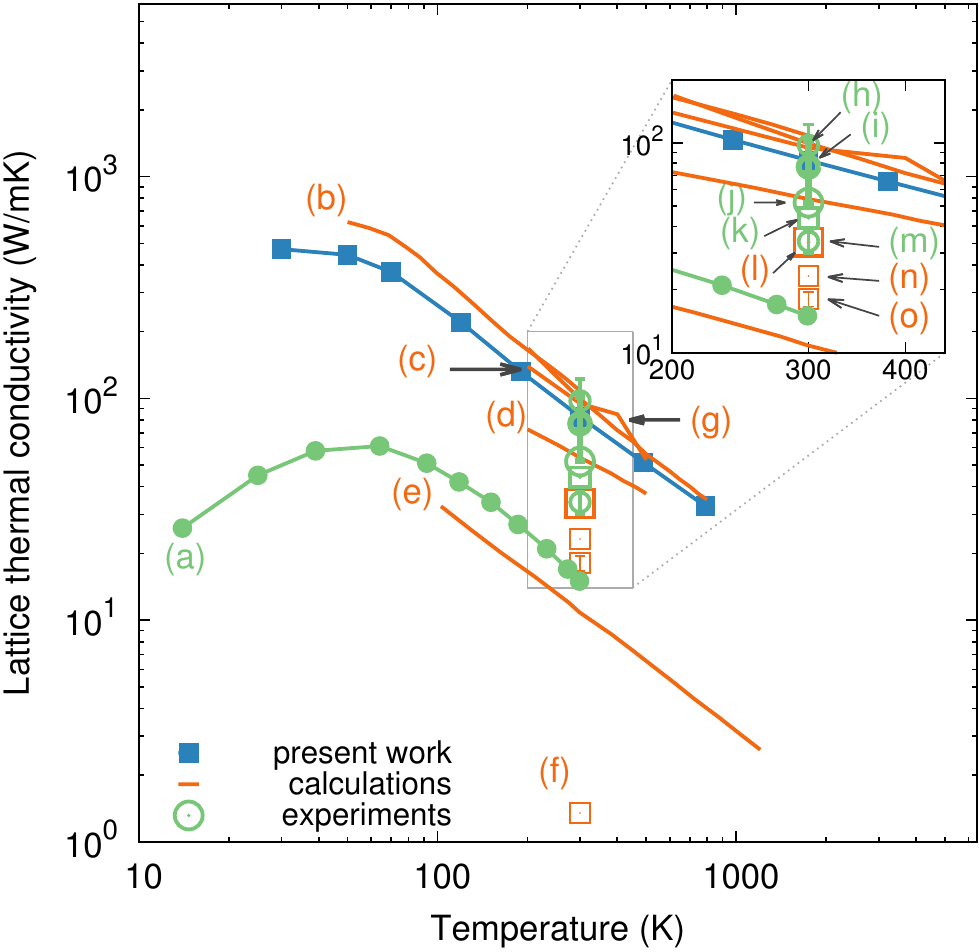}
  \caption{
    Experimental and theoretical results for the in-plane (basal plane) thermal conductivity of MoS$_2$. Calculations and experiments are from (a) \cite{PisJacBar15}, (b) \cite{LiCarMin13}, (c) \cite{PenZhaSha16a}, (d) \cite{GuYan14}, (e) \cite{DinCheXia16}, (f) \cite{LiuZhaPei13}, (g) \cite{KanYapKin16}, (h) \cite{LiuChoCah14}, (i) \cite{ZhaSunLi15}, (j) \cite{SahGauAhm13}, (k) \cite{MurVarGen13}, (l) \cite{PenZhaSha16b}, (m) \cite{YanSimBer14}, (n) \cite{CaiLanZha14}, (o) \cite{VarPatMur10}.
    Data from Refs.~\onlinecite{LiCarMin13, PenZhaSha16a, GuYan14, LiuZhaPei13, KanYapKin16, ZhaSunLi15, PenZhaSha16b, YanSimBer14, CaiLanZha14} was obtained for monolayers; the data from Ref.~\onlinecite{SahGauAhm13} is for few-layer systems.
  }
  \label{fig:conductivity-MoS2}
\end{figure}

Unfortunately, values for the thermal conductivities reported in the literature exhibit a wide spread. For example in the case of nominally single-crystalline MoS$_2$, experimental values for the in-plane (basal plane) thermal lattice conductivity vary over one order of magnitude ranging from around \unit[20]{W/K\,m} \cite{PisJacBar15} up to \unit[110]{W/K\,m} \cite{LiuChoCah14} at room temperature (\fig{fig:conductivity-MoS2}). This can be partly attributed to the challenges associated with experimental measurements of the thermal conductivity in nanostructures with pronounced anisotropy, see e.g., Refs.~\onlinecite{WilCah14, LiuChoCah14}. Possibly even more crucial are defects and sample size effects, as the growth of large high-quality TMD single crystals is very time consuming \cite{LiuChoCah14}. The extreme sensitivity to structure has been possibly most impressively demonstrated in the case of WSe$_2$ \cite{ChiCahNgu07, NguBerLin10}, for which the out-of-plane (through plane) thermal conductivity $\kappa_\perp$ has been shown to vary by almost two orders of magnitude at room temperature. This variation can in fact be rationalized in terms of the microstructure, in particular planar defects such as stacking faults and subtle variations in layer spacing \cite{ErhHylLin15}.

Similar to the experimental data, calculated values for the thermal conductivity cover a wide range as well. \textit{Ab-initio} calculations based on Boltzmann transport theory  in combination with density functional theory have only become available relatively recently \cite{LinBroMin10, EsfCheSto11, TiaGarEsf12, LiCarKat14, TogChaTan15}. Still, as illustrated by the case of MoS$_2$ (\fig{fig:conductivity-MoS2}), calculations have usually been restricted to monolayers \cite{LiCarMin13, PenZhaSha16a, GuYan14, LiuZhaPei13, KanYapKin16, PenZhaSha16b, CaiLanZha14}. This is at least in part due to the fact that computational studies of bulk systems \cite{VarPatMur10, DinCheXia16} require taking into account the vdW forces that mediate interlayer binding. These interactions are, however, not captured by common semi-local exchange-correlation (XC) functionals \cite{BerHyl14b}, including widely popular functionals such as PBE \cite{PerBurErn96} and PBEsol \cite{PerRuzCso08}. In some cases this shortcoming has been addressed by using semi-empirical methods \cite{GanSch14}. As will be shown below, in general, the structural parameters of TMDs as well as other quantities that affect the thermal conductivity are, however, very sensitive to the treatment of exchange and correlation. Furthermore, since vdW forces are rather weak and computational noise can blur anharmonic effects, both the choice of the XC functional and the convergence of the computational parameters require special care.

This perspective motivates the present study, in which we have carefully evaluated both the in-plane and out-of-plane thermal conductivities of Mo and W-based TMDs. To this end, we employ a combination of density functional and Boltzmann transport theory calculations based on the vdW density functional method \cite{LanLunCha09} in combination with a recently formulated consistent-exchange part \cite{BerHyl14b, BerArtCoo14}, which has already been found to work very well for e.g., WSe$_2$ \cite{ErhHylLin15}. In the following, we first demonstrate that this approach yields an excellent description of the structural parameters of Mo and W-based TMDs at finite temperatures. We then carefully assess the relevant computational parameters before conducting a comprehensive investigation of the thermal conductivities. Since the largest contribution to the thermal conductivities stems from modes with mean free paths (phonon-phonon scattering limited) of more than $1\mu\text{m}$, both in-plane and out-of-plane conductivities are in practice often limited by structural incoherence. The thermal conductivities $\kappa$ are found to increase from MTe$_2$ to MS$_2$ but, in opposition to the trend expected based on the group velocities, $\kappa$ tends to be higher for WS$_2$ than for the respective Mo compound. This behavior is shown to be due to higher lifetimes in the former case, which can be rationalized in terms of the scattering condition and the different phononic band gaps.

\section{Methodology}

\subsection{Thermal conductivity}

In general the thermal conductivity comprises both an electronic $\kappa_e$ and a phononic (lattice) part $\kappa_l$. According to the Wiedemann-Franz law the electronic contribution $\kappa_e$ is closely related to the electrical conductivity. Since the TMDs of interest in the present work have comparably large band gaps $\kappa_e$ is usually much smaller than $\kappa_l$. For example in the case of the in-plane conductivity in MoS$_2$ $\kappa_e$ reaches only about 5\%\ of the value of $\kappa_l$ at room temperature \cite{PisJacBar15} and the ratio is even smaller below 300\,K. In the present work, we therefore focus entirely on the lattice contribution $\kappa_l$ and from here on drop the subscript $l$.

To calculate the lattice thermal conductivity we utilize Boltzmann transport theory within the relaxation time approximation. In this approximation each mode $\lambda=(\vec{q},p)$, where $\vec{q}$ is the phonon wave-vector and $p$ is the band index, is associated with a relaxation time $\tau_\lambda$. The total relaxation time is the result of several scattering processes, and in the present work we consider phonon-phonon scattering as well as isotopic and boundary scattering. If one assumes that each scattering rate individually contributes in parallel, the total relaxation time for a phonon mode is given by Matthiessen's rule,
\begin{align} 
  \tau_{\lambda}^{-1}
  &= \tau_{\text{ph-ph},\lambda}^{-1}
  +  \tau_{\text{iso},\lambda}^{-1}
  +  \tau_{\text{boundary},\lambda}^{-1}.
  \label{eq:matthiesen-rule}
\end{align}

\paragraph*{Isotopic scattering} is the result of variations in the atomic masses due to the natural isotope distribution. The corresponding relaxation time contribution $\tau_{\text{iso},\lambda}$ has been calculated according to second-order perturbation theory \cite{Tam83} using isotope distributions and masses from Ref.~\onlinecite{LaeBohBiv03}.

\paragraph*{Boundary scattering} is accounted for by assuming that the mean free path (MFP) of any phonon mode is capped by an intrinsic length scale $L$, which in the most simple case corresponds to the sample size \cite{Zim60},
\begin{align}
  \tau_{\text{boundary},\lambda}^{-1} = v_{\lambda}/L.
  \label{eq:boundary-scattering}
\end{align}
This expression represents the limit, in which the scattering event is fully diffusive, equivalent to a vanishing specularity parameter \cite{CheTieWu94, NikPokAsk09}. Below we will treat this model as a means to establish the characteristic length scale $L$ that is representative of the (temperature independent) structural homogeneity of the material. We note that the model was used in a similar fashion in Ref.~\onlinecite{KatTogTan15} to describe the effect of nanostructuring in Zn chalcogenides.

\paragraph*{Phonon-phonon scattering} is computationally the most intricate contribution. The corresponding lifetime $\tau_{\text{ph-ph},\lambda}$ can be obtained using perturbation theory on top of a harmonic description of lattice vibrations. The phonon-phonon limited lifetime is then obtained as the inverse of the self energy $\tau_{\text{ph-ph},\lambda} = 1 / 2\Gamma_{\lambda}(\omega_\lambda)$, where the self-energy is given by \cite{TogChaTan15}
\begin{align}
  \begin{split}
    &\Gamma_\lambda(\omega)
    = \frac{18\pi}{\hbar^2}\sum_{\lambda'\lambda''}| \Phi_{-\lambda\lambda'\lambda''}|^2
    \Big \{
    \\
    &(n_{\lambda'} + n_{\lambda''} + 1)\;\delta(\omega -\omega_{\lambda'} -\omega_{\lambda''})
    \\
    &+ (n_{\lambda'}-n_{\lambda''})\big ( \delta(\omega + \omega_{\lambda'} - \omega_{\lambda''}) - \delta(\omega - \omega_{\lambda'} + \omega_{\lambda''})\big )
    \Big \}.
  \end{split}
  \label{eq:self-energy}
\end{align}
Here, $\Phi_{-\lambda\lambda'\lambda''}$ is obtained from the third order interatomic force constant (IFC) matrix and $n_{\lambda}$ is the Bose-Einstein distribution. The mode frequencies $\omega_\lambda$ can be obtained in the usual fashion from the second order IFCs \cite{Zim60}.

Phonon scattering processes must obey (\emph{i}) momentum conservation, $\vec{q}_\lambda + \vec{q}_{\lambda'} + \vec{q}_{\lambda''} + \vec{G} = \vec{0}$, where $\vec{G}$ is a reciprocal lattice vector, and (\emph{ii}) energy conservation, $\delta(\omega_\lambda \pm \omega_{\lambda'} \pm \omega_{\lambda''})$, where the signs are determined by the type of scattering event. Condition (\emph{i}) is included in the constructing of the third-order IFCs while condition  (\emph{ii}) is apparent in Eq.~\eqref{eq:self-energy}. The structure of the self-energy and accordingly the lifetimes is thus determined to a large extent by the geometry of the Brillouin zone and the phonon dispersion \cite{Zim60}. This observation allows one to identify general trends in the lifetime spectrum already on the basis of the phonon dispersion and thus the second-order IFCs. In this context the weighted joint density of states introduced by Togo \textit{et al.} \cite{TogChaTan15} is a very useful quantity. Specifically, we considered the weighted joint density of states for so-called class 1 processes \cite{Zim60}, which correspond to collisions by which two phonons combine to form one phonon. It is defined as \footnote{
  For the analysis presented in \fig{fig:conductivity-WS2-WSe2} class 2 processes (decay of one phonon into two phonons) are less relevant and therefore not considered further.
}
\begin{align}
  \begin{split}
    N_2^{(1)}(\vec{q},\omega)
    =&
    \dfrac{1}{N}
    \sum_{\lambda',\lambda''} \Delta\left( -\vec{q} + \vec{q}' + \vec{q}'' \right) \left( n_{\lambda'} - n_{\lambda''} \right) \\
    &\times \left[ \delta \left( \omega + \omega_{\lambda'} - \omega_{\lambda''} \right)
      -  \delta \left( \omega - \omega_{\lambda'} + \omega_{\lambda''} \right) \right],
  \end{split}
  \label{eq:wjdos}
\end{align}
where $N$ is the number of unit cells in the crystal and $\Delta( -\vec{q} + \vec{q}' + \vec{q}'' )$ embodies the momentum conservation condition expressed above. $N_2^{(1)}(\vec{q},\omega)$ thus effectively counts the number of collision processes that contribute to the phonon-phonon scattering time of a given mode. By comparison with the full expression one recognizes as the main difference the occurrence of third-order derivatives of the total energy in Eq.\eqref{eq:self-energy}\footnote{
  Also compare Eqs.~(3.2.11-12) in Ref.~\onlinecite{Zim60} and Eq.~(1) in Ref.~\onlinecite{TogChaTan15}.
}, which represent the efficiency of the scattering processes that are energy and momentum allowed. By contrast, Eq.~\eqref{eq:wjdos} requires only knowledge of the second-order force constants.

Finally, the \emph{full lattice thermal conductivity tensor} is obtained by summing over all modes \cite{Sri90}
\begin{align}
  \vec{\kappa}(T) = 
  \dfrac{1}{N_{\vec{q}}\Omega}
  \sum_\lambda
  \underbrace{\tau_\lambda(T) \vec{v}_\lambda}_{\textstyle\vec{\Lambda}_\lambda(T)} \otimes \vec{v}_\lambda c_\lambda(T).
  \label{eq:full-thermal-conductivity}
\end{align}
Here $\Omega$ is the unit cell volume, $N_{\vec{q}}$ denotes the number of $q$-points, $\vec{v}_\lambda=\nabla \omega_\lambda$ is the group velocity, $\vec{\Lambda}_\lambda$ is the phonon MFP, and $c_\lambda(T)$ is the mode specific heat capacity. For analyzing, e.g., the sensitivity of the thermal conductivity to structural inhomogeneities it is convenient to consider the cumulative thermal conductivity, which is given by
\begin{align}
  \widebar{\vec{\kappa}}(\Lambda) &=
  \dfrac{1}{N_{\vec{q}}\Omega}
  \sum_\lambda^{\Lambda_\lambda<\Lambda}
  \vec{\Lambda}_\lambda(T) \otimes \vec{v}_\lambda c_\lambda(T).
  \label{eq:cumulative-thermal-conductivity}
\end{align}
If the MFP is uniformly limited to a constant value $\widetilde{\Lambda}$, one obtains the so-called small-grain conductivity \cite{LiCarKat14}, which is given by
\begin{align}
  \kappa_\text{sg} &=
  \frac{1}{N_{\vec{q}}\Omega}
  \sum_{\lambda}
  v_\lambda c_\lambda(T).
  \label{eq:small-grain-conductivity}
\end{align}
The small-grain conductivity represents the limit, in which scattering is dominated by an intrinsic length scale as for example in the case of nanostructuring.

\subsection{Computational details}

Density functional theory calculations were carried out using the projector augmented wave method \cite{Blo94, KreJou99} as implemented in the Vienna ab-initio simulation package (\textsc{vasp}) \cite{KreHaf93, KreFur96b}. To assess the sensitivity of our results to the treatment of exchange-correlation effects we used both the local density approximation (LDA) and the van der Waals density functional (vdW-DF) method that captures non-local correlations \cite{RydDioJac03, DioRydSch04, ThoCooLi07, BerCooLee15}. With regard to the latter, we considered both the empirically adjusted PBE exchange part from Ref.~\onlinecite{KliBowMic11} (vdW-DF-optPBE) and the recently developed consistent exchange version (vdW-DF-CX) \cite{BerHyl14b, BerArtCoo14} as implemented in \textsc{vasp} \cite{KliBowMic11, Bjo14}. The plane wave energy cutoff energy was set to \unit[290]{eV} in the calculations of WSe$_2$, MoSe$_2$, WTe$_2$ and MoTe$_2$ and to \unit[336]{eV} in the calculations of WS$_2$ och MoS$_2$. In calculations based on the primitive cell the Brillouin zone was sampled using a $\Gamma$-centered $12\times12\times3$ $\vec{k}$-point mesh.

Thermal conductivities and other phonon related quantities where obtained with the \textsc{phonopy} \cite{TogObaTan08, TogTan15} and \textsc{phono3py} \cite{TogChaTan15} packages. The convergence of the lattice thermal conductivity with respect to $\vec{q}$-point sampling mesh, displacement amplitude, supercell size as well as the cutoff for the maximal range of force interactions was analyzed as described in \sect{sect:convergence-of-thermal-conductivity} below. The final calculations for both second and third order force constants were conducted using supercells comprising $3\times3\times1$ primitive unit cells while a $\Gamma$-centered $4\times4\times3$ grid was utilized for $\vec{k}$-point sampling. The displacement amplitude employed in the calculation of finite differences was set to \unit[0.09]{\r{A}}. This value was obtained by balancing the need to reduce the numerical noise in the computation of soft interlayer force components while remaining in the harmonic (linear response) regime. For computational efficiency forces were only computed for pairs and triplets within a cutoff range of \unit[3.8]{\AA}; this includes interactions up to the third nearest neighbor shell for in-plane terms and between neighboring layers in the out-of-plane direction for all considered materials. For the lattice thermal conductivity calculations a tetrahedron method was used for Brillouin zone integrations while employing a $21\times 21\times 13$ $\vec{q}$-point mesh.

The structural properties at finite temperature were obtained at the level of the quasi-harmonic approximation as implemented in \textsc{phonopy} \cite{TogObaTan08, TogTan15}. To this end, the second order IFCs were computed at seven different volumes between 95 and 105\%\ of the respective equilibrium volume.

\section{Results and discussion}

\subsection{Description of van der Waals solids}
\label{sect:description-of-vdW-solids}

\begin{figure}[b]
\includegraphics[width=0.85\linewidth]{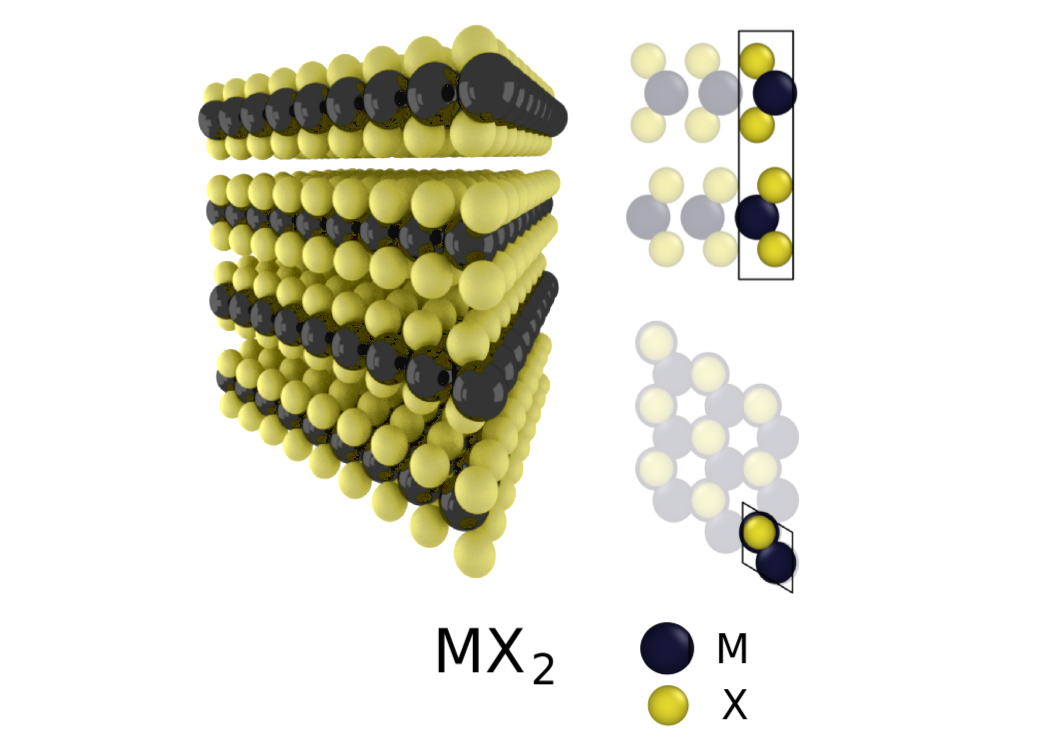}
  \caption{
    Crystal structure of molybdenum and tungsten based transition metal dichalcogenides (spacegroup $P6_3/mmc$, ITC no. 194). The transition metal and chalcogenide species correspond to M and X, respectively.
    Structures were created using the atomic simulation environment \cite{BahJac02} and visualized with \textsc{ovito} \cite{Stu10} as well as Blender \cite{blender}.
  }
  \label{fig:structure}
\end{figure}

\begin{table}
  \centering
  \caption{
    Comparison of structural parameters for WSe$_2$ from experiment \cite{SchDeBJel87,Kal83} and calculation.
    $a$, $c$; in-plane and out-of-plane lattice constants (\AA);
    $z_\text{Se}$: internal parameter, which specifies the position of the Se atoms.
  }
  \label{tab:structural-parameters-WSe2}
  \begin{tabularx}{\columnwidth}{lX*{3}dX*{2}d}
    \toprule
    &
    & \multicolumn{3}{c}{Calculations}
    &
    & \multicolumn{2}{c}{Experiment}
    \\

    \cline{3-5}
    \cline{7-8}
    \\[-9pt]
    
    &
    & \multicolumn{1}{c}{LDA}
    & \multicolumn{1}{c}{vdW-optPBE}
    & \multicolumn{1}{c}{vdW-CX}
    \\[3pt]
    
    \multicolumn{6}{l}{zero K excluding zero-point vibrations} \\
    & $a$     &  3.250 &  3.341 &  3.277 & \\
    & $c$     & 12.819 & 13.550 & 12.942 & \\
    & $z_\Se$ &  0.620 &  0.626 &  0.620 & \\[6pt]
    
    \multicolumn{6}{l}{zero K with zero-point vibrations} \\
    & $a$     &  3.250 &  3.339 &  3.279 \\
    & $c$     & 12.824 & 13.500 & 12.991 \\
    & $z_\Se$ &  0.620 &  0.625 &  0.620 \\[6pt]

    \multicolumn{6}{l}{300\,K} \\
    & $a$     &  3.250 &  3.339 &  3.279 &&  3.282 &  3.286 \\
    & $c$     & 12.832 & 13.508 & 12.998 && 12.960 & 12.960 \\
    & $z_\Se$ &  0.620 &  0.625 &  0.621 &&  0.621 &  0.621 \\

    \botrule
  \end{tabularx}
\end{table}

\begin{figure}
  \centering
\includegraphics[width=0.88\columnwidth]{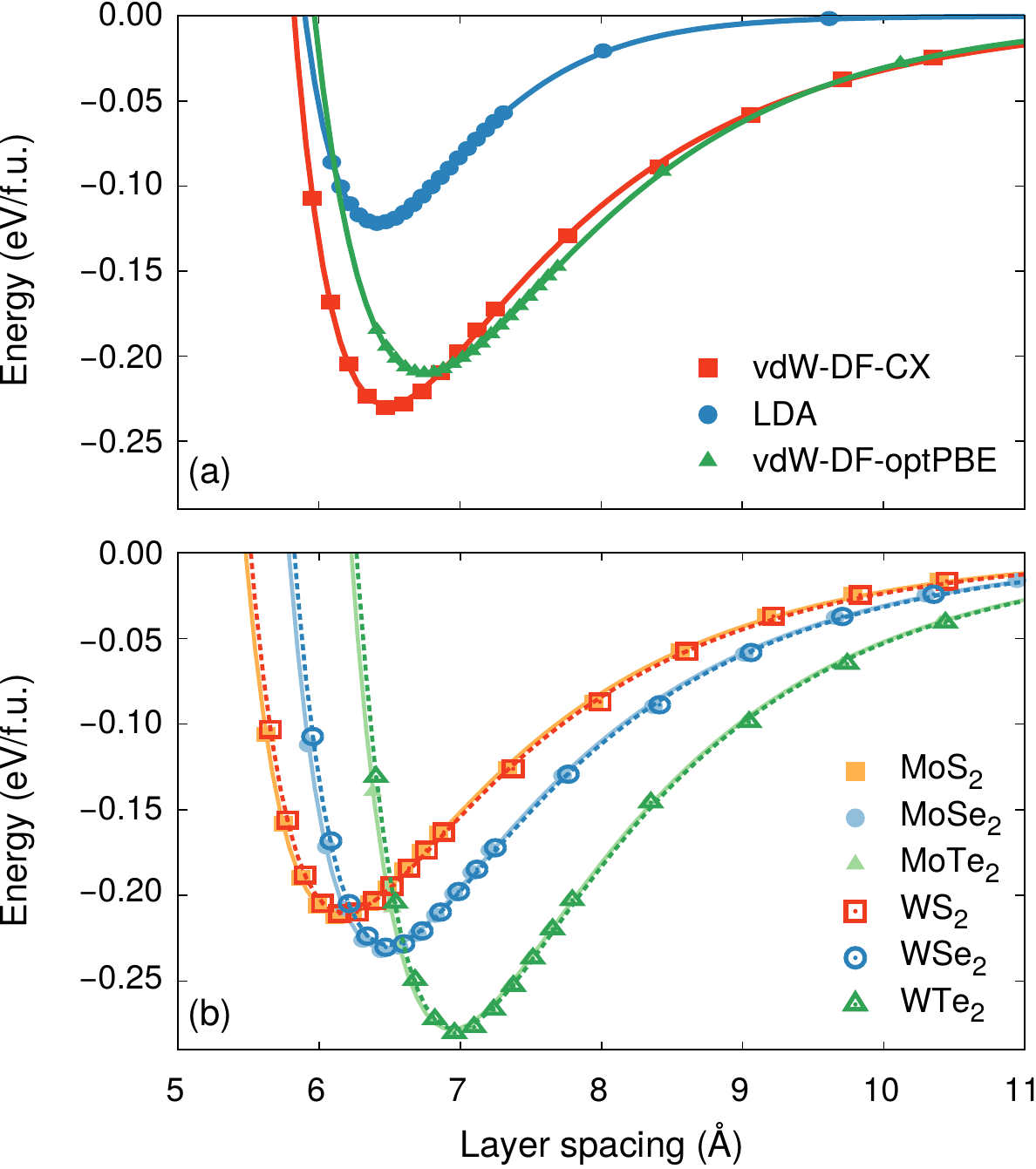}
  \caption{
    Binding energy as a function of interlayer spacing (a) for WSe$_2$ as obtained using different XC functionals and (b) for Mo and W-based sulfides, selenides, and tellurides calculated using the vdW-DF-CX functional.
    The in-plane lattice constant was held fixed at the equilibrium value.
  }
  \label{fig:energy-separation}
\end{figure}

\begin{table}
  \centering
  \caption{
    Comparison of structural parameters for di-sulfides, selenides, and tellurides of Mo and W in spacegroup $P6_3/mmc$ (ITC no. 194) as obtained from calculations using the vdW-DF-CX functional and experiment. Results from calculations that  exclude of zero-point vibrations are shown in brackets.
    The in-plane and out-of-plane lattice constants $a$ and $c$ are given in units of {\AA}ngstr\"om.
  }
  \label{tab:structural-parameters-TMDs}
  \begin{tabularx}{\columnwidth}{*{2}X*{7}d}
    \toprule
    \multicolumn{2}{l}{Material}
    & \multicolumn{5}{c}{Calculation}
    & \multicolumn{2}{c}{Experiment}
    \\

    \colrule
    \\[-9pt]

    &
    & \multicolumn{2}{c}{0\,K}
    && \multicolumn{1}{c}{300\,K}
    && \multicolumn{2}{c}{300\,K}
    \\

    \cline{3-4}
    \cline{6-6}
    \cline{8-9}
    \\[-6pt]
    
    \multicolumn{9}{l}{MoS$_2$, Refs.~\onlinecite{Kal83,BroDeBJel86}} \\
    & $a$     &  3.152 &  (3.149) &&  3.152 &&  3.160 &  3.160 \\
    & $c$     & 12.291 & (12.225) && 12.295 && 12.294 & 12.290 \\
    & $z_\S$  &  0.622 &  (0.621) &&  0.622 &&  0.621 &  0.620 \\[6pt]
    \multicolumn{9}{l}{MoSe$_2$, Refs.~\onlinecite{Kal83,BroDeBJel86}} \\
    & $a$     &  3.280 &  (3.278) &&  3.280 &&  3.289 &  3.288 \\
    & $c$     & 12.920 & (12.875) && 12.928 && 12.927 & 12.930 \\
    & $z_\Se$ &  0.620 &  (0.621) &&  0.621 &&  0.621 &  0.620 \\[6pt]
    \multicolumn{9}{l}{MoTe$_2$, Refs.~\onlinecite{PuoNew61,KnoMac61}} \\
    & $a$     &  3.504 &  (3.501) &&  3.504 &&  3.519 &  3.518 \\
    & $c$     & 13.904 & (13.865) && 13.913 && 13.964 & 13.974 \\
    & $z_\Te$ &  0.619 &  (0.619) &&  0.619 &&  0.625 &  0.621 \\[6pt]
    \multicolumn{9}{l}{WS$_2$, Refs.~\onlinecite{SchDeBJel87,Kal83}} \\
    & $a$     &  3.152 &  (3.150) &&  3.152 &&  3.153 &  3.154 \\
    & $c$     & 12.358 & (12.288) && 12.365 && 12.323 & 12.360 \\
    & $z_\S$  &  0.623 &  (0.622) &&  0.623 &&  0.623 &  0.614 \\[6pt]
    \multicolumn{9}{l}{WSe$_2$, Refs.~\onlinecite{SchDeBJel87,Kal83}} \\
    & $a$     &  3.279 &  (3.277) &&  3.279 &&  3.282 &  3.286 \\
    & $c$     & 12.991 & (12.942) && 12.998 && 12.960 & 12.980 \\
    & $z_\Se$ &  0.620 &  (0.620) &&  0.621 &&  0.621 &  0.620 \\[6pt]
    \multicolumn{9}{l}{WTe$_2$} \\
    & $a$     &  3.506 &  (3.503) &&  3.506 \\
    & $c$     & 13.954 & (13.916) && 13.961 \\
    & $z_\Te$ &  0.620 &  (0.619) &&  0.620 \\
    \botrule
  \end{tabularx}
\end{table}

\paragraph*{Tungsten diselenide.}
Molybdenum and tungsten based transition metal dichalcogenides (TMDs) are among the most widely investigated vdW solids. They adopt layered structures with stoichiometry MX$_2$ (M=Mo, W; X=S, Se, Te) that are composed of two-dimensional sheets with strong intralayer bonding coupled to each other via comparably weak vdW interactions. With the exception of WTe$_2$ the equilibrium structures belong to spacegroup $P6_3/mmc$ (International Tables of Crystallography no. 194, see \fig{fig:structure}). In equilibrium WTe$_2$ adopts an orthorhombic crystal structure that belongs to spacegroup $Pmn2_1$ (ITC no. 31) \cite{Bro65}. It is included here in spacegroup $P6_3/mmc$ to exhibit chemical trends and since it be incorporated in multilayer vdW solids with hexagonal symmetry.

\begin{figure*}
  \centering
\includegraphics[scale=\myscale]{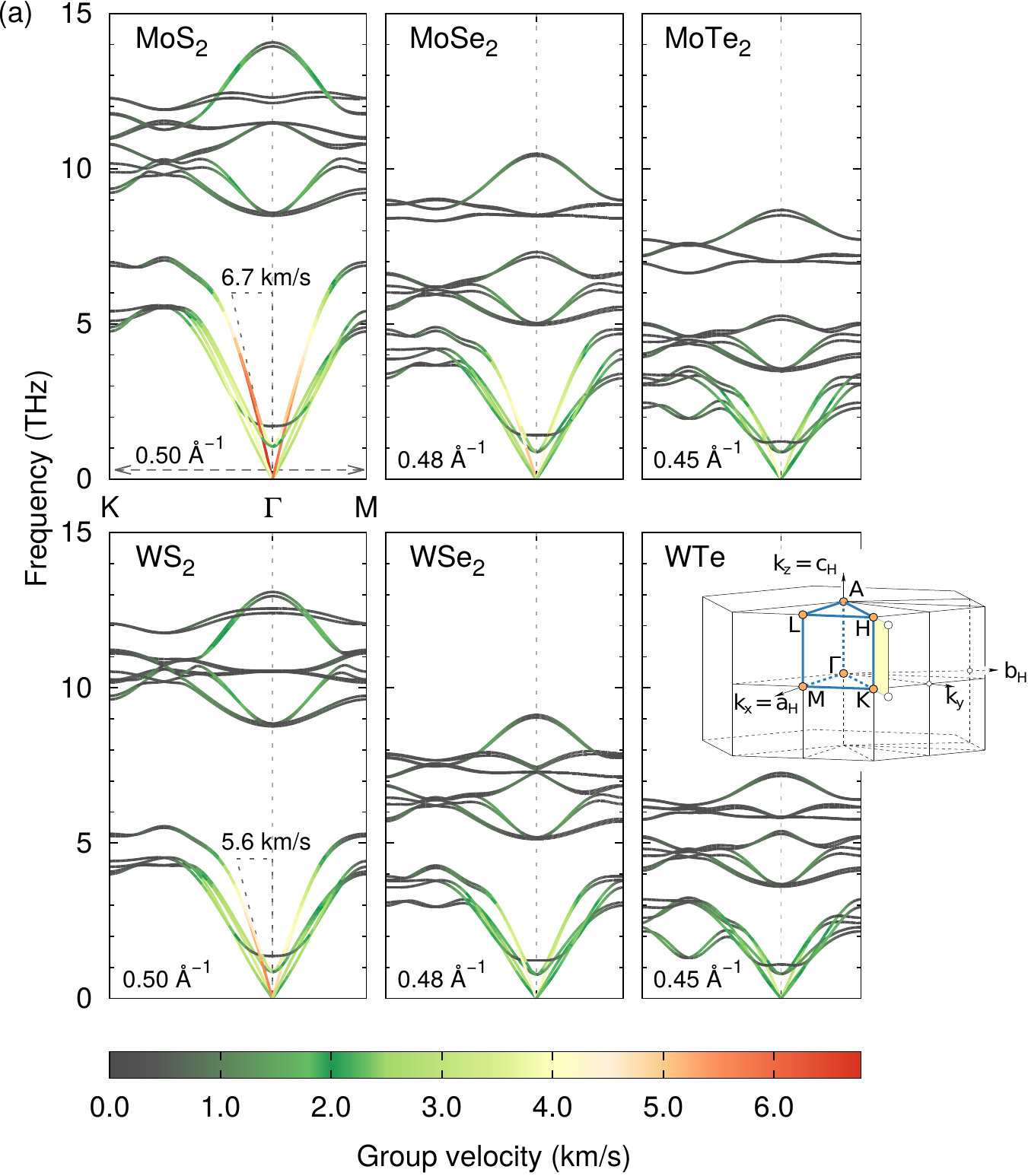}
\includegraphics[scale=\myscale]{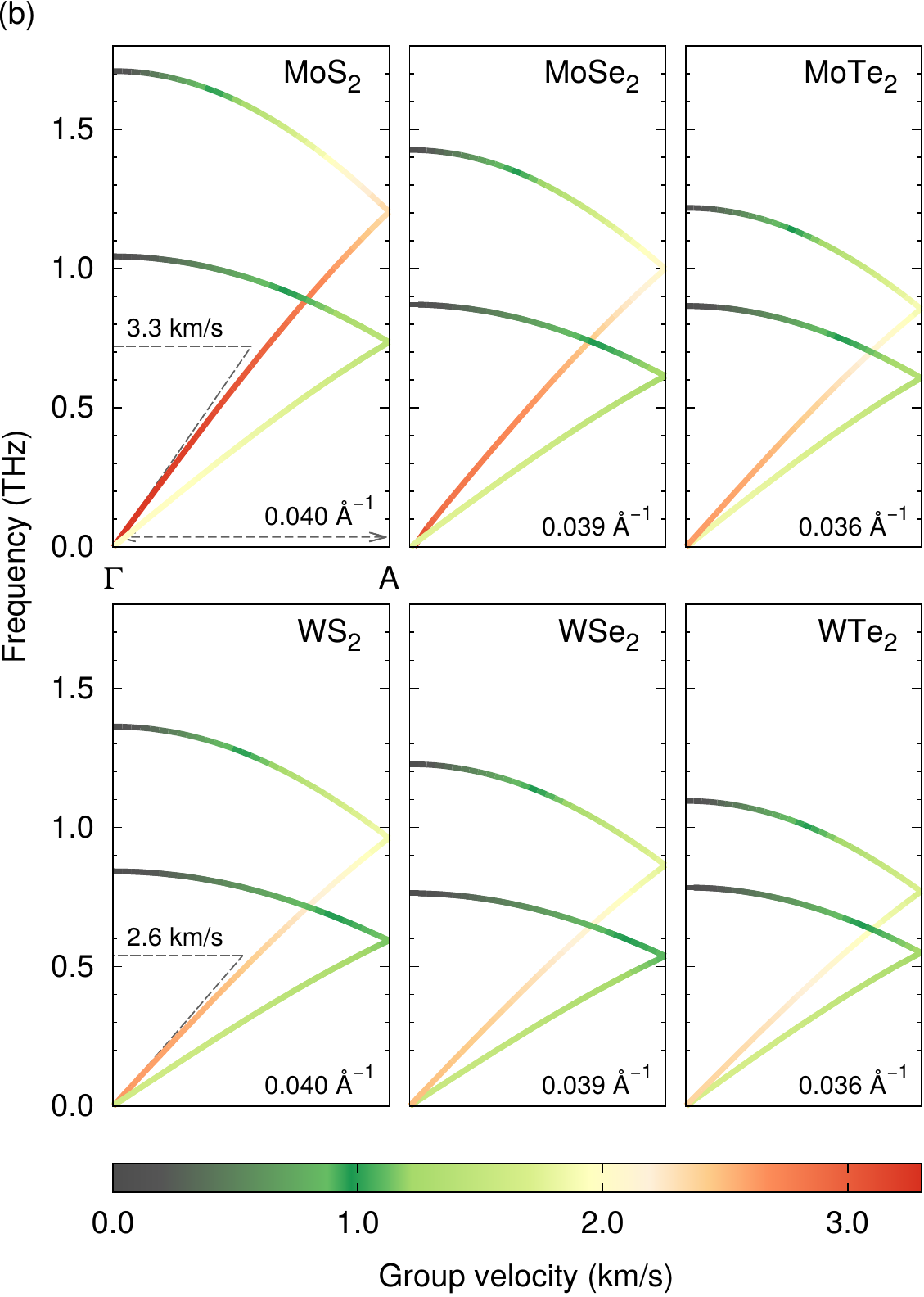}
  \caption{
    (a) In-plane and (b) out-of-plane phonon dispersion for Mo and W-based TMDs color coded by group velocity. The inset in (a) shows the Brillouin zone for spacegroup $P6_3/mmc$ (ITC no. 194) \cite{TasFloOro12}.
    The number in the left bottom corner of plot represents the dimension of the Brillouin zone along the direction shown by the dashed arrows.
    The longitudinal sound velocities along (a) $\Gamma$-X and (b) $\Gamma$-A are indicated by dashed triangles for each material.
  }
  \label{fig:phonon-dispersions}
\end{figure*}

For WSe$_2$ the structural parameters at 300\,K were computed using the local density approximation (LDA) as well as the vdW-DF-optPBE and vdW-DF-CX functionals [\tab{tab:structural-parameters-WSe2} and \fig{fig:energy-separation}(a)]. We also considered the PBE functional but the lack of vdW binding gives rise to extremely weak interlayer binding and a very poor description of the structure, in particular the out-of-plane lattice parameter.

The closest agreement with the structural reference data is obtained for the vdW-DF-CX functional, which yields values for the in-plane and out-of-plane lattice parameters that are within respectively 0.1\%\ and 0.3\%\ of the experimental data. We are not aware of higher-level (experiment or calculation) reference data for the interlayer binding energy [\fig{fig:energy-separation}(a)] but note that the vdW-DF-CX functional has been shown to yield excellent binding energies for other vdW bonded systems \cite{BerCooLee15}.

The vdW-DF-optPBE functional was obtained in semi-empirical fashion by combining the non-local vdW-DF correlation with the rescaled exchange part of the PBE functional \cite{PerBurErn96, KliBowMic11}. Here, it is found to overestimate both in-plane (1.7\%) and out-of-plane (4.2\%) lattice constants of WSe$_2$ notably; it also yields a slightly smaller value for the interlayer cohesion than the vdW-DF-CX functional.

The LDA results for both lattice constants are within 1\%\ of the experimental values. This result is partially surprising in so far as the LDA actually does not account for dispersive vdW interactions, and the good agreement is rather the result of the characteristic LDA overbinding, which has been pointed out previously \cite{RydDioJac03, MurLeeLan09}. The LDA thus yields the correct result for the wrong reasons \cite{Gul12}, which becomes more evident when considering the binding energy curve [\fig{fig:energy-separation}(a)]. The asymptotic behavior of the LDA data clearly differs from the two vdW functionals and yields only about half of the interlayer binding energy. The energy landscape around the equilibrium spacing is, however, similar to the one obtained with the vdW-DF-CX functional.

\paragraph*{Extension to other TMDs.}
Based on the results for WSe$_2$ we only considered the vdW-DF-CX functional for the analysis of the other Mo and W-based TMDs. This functional generally achieves very good agreement with experimental measurements (\tab{tab:structural-parameters-TMDs}) as the deviations from the reference data generally do not exceed 0.4\%\ and are on average below 0.2\%.

The results show the structural parameters are barely affected by the transition metal, while the chalcogenide species has a very strong effect as the lattice parameters increase in the order S--Se--Te. As will be discussed in more detail below, this has a direct impact on the vibrational properties as the size of the Brillouin zone is inversely proportional to the lattice parameters (see \fig{fig:phonon-dispersions}).

\subsection{Convergence of the thermal conductivity}
\label{sect:convergence-of-thermal-conductivity}

\begin{figure*}
  \centering
\includegraphics[width=0.98\linewidth]{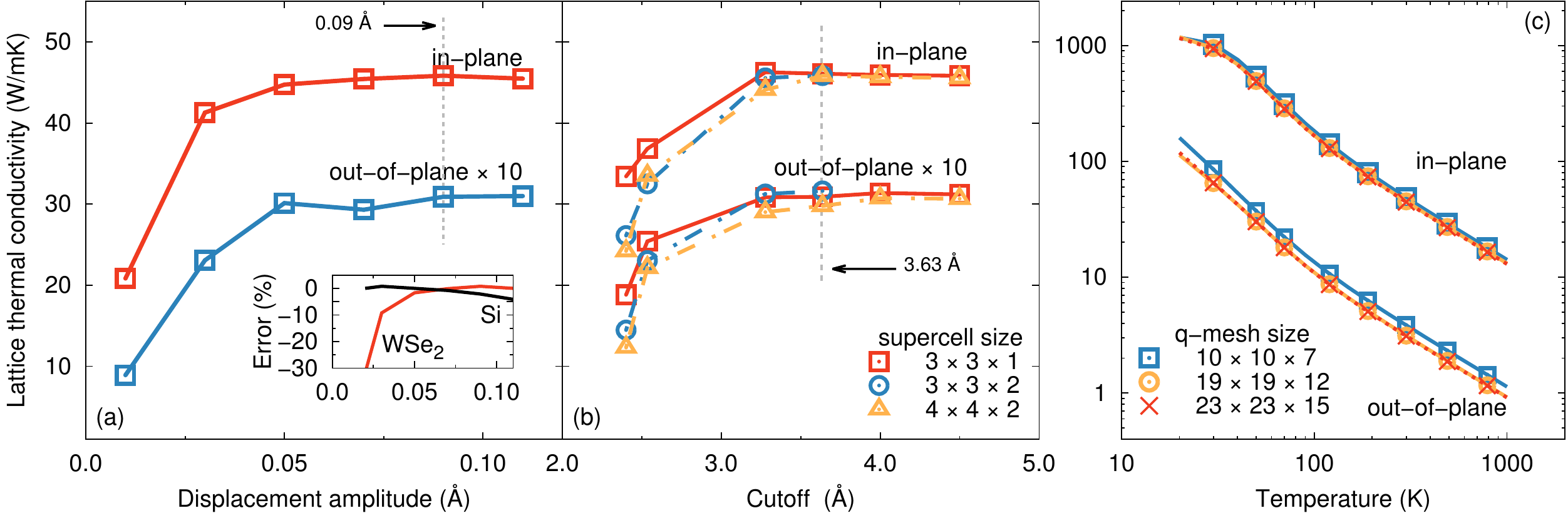}
  \caption{
    Convergence of thermal conductivity with respect to different numerical parameters.
    (a) Calculated lattice thermal conductivity in WSe$_2$ at \unit[300]{K} as a function of the displacement amplitude used for calculating the IFCs. The out-of-plane conductivity has been scaled by 10 for clarity.
    The inset compares the relative error in the in-plane conductivity for WSe$_2$ with the case of silicon, which is a purely covalently bonded material and much less sensitive to the choice of the displacement amplitude.
    (b) Convergence of $\kappa$ at \unit[300]{K} with respect to supercell size and interaction cutoff distance.
    (c) Convergence with respect to the $\vec{q}$-point mesh used for Brillouin zone integrations.
  }
  \label{fig:convergence}
\end{figure*}

Since the vdW forces acting between layers are much weaker than the covalent and ionic interactions in denser materials, they are more prone to numerical errors. This is partially compensated by using tight convergence parameters e.g., for the plane wave cutoff energy and the termination of the electronic self-consistency loop. When calculating second and especially third-order derivatives using finite differences errors in the forces are, however, enhanced. We therefore carefully tested the effect of the displacement amplitude $\Delta r$ used for computing the IFCs on the calculated lattice thermal conductivity.

The thermal conductivity is in fact very sensitive to the displacement amplitude $\Delta r$ [\fig{fig:convergence}(a)]. While in the  case of silicon [inset in \fig{fig:convergence}(a)] $\kappa$ is only weakly dependent on $\Delta r$, for WSe$_2$ the thermal conductivity is dramatically underestimated for smaller values of $\Delta r$.
Since one usually strives to use small values for $\Delta r$ in order to remain in the linear response regime, common (default) values for $\Delta r$ typically fall in the range between 0.01 and 0.03\,\AA\ \cite{LiCarKat14, TogChaTan15}. In the case of WSe$_2$ these values cause a pronounced error in $\kappa$, as $\Delta r$ values $\gtrsim0.05\,\text{\AA}$ are required to obtain convergence. We therefore adopted a value of \unit[0.09]{\AA} for the bulk of our calculations.

The calculation of the thermal conductivity is also affected by supercell size and the cutoff imposed on the interaction range. Based on the results of our convergence study [\fig{fig:convergence}(b)], production runs were conducted using supercells comprising $3\times 3\times 1$ unit cells and interactions were included up to the third neighbor shell  in-plane and the first neighbor shell out-of-plane (equivalent to a cutoff of \unit[3.63]{\AA} in the case of WSe$_2$).

Finally, the thermal conductivity is affected by the density of the $\vec{q}$-point grid used for Brillouin zone integrations. In this regard, we find that a $19\times 19\times 12$ $\vec{q}$-point mesh corresponding to approximately 4300 $\vec{q}$-points in the full Brillouin zone achieves a convergence level that is comparable to the other parameters considered here [\fig{fig:convergence} (c)].

\subsection{Thermal conductivity in WS$_2$ and WSe$_2$}

Having established the quality of the underlying XC functional with regard to structural parameters (\sect{sect:description-of-vdW-solids}) as well as the numerical convergence of our calculations (\sect{sect:convergence-of-thermal-conductivity}), we can now compare the calculated thermal conductivities with experiment. To this end, we first consider WS$_2$ and WSe$_2$, for which experimental data over a wide temperature range is available for both the in-plane and out-of-plane conductivities of nominally single-crystalline material \cite{PisJacGaa16, ChiCahNgu07}.

\begin{figure}[b]
 \centering
\includegraphics[width=\linewidth]{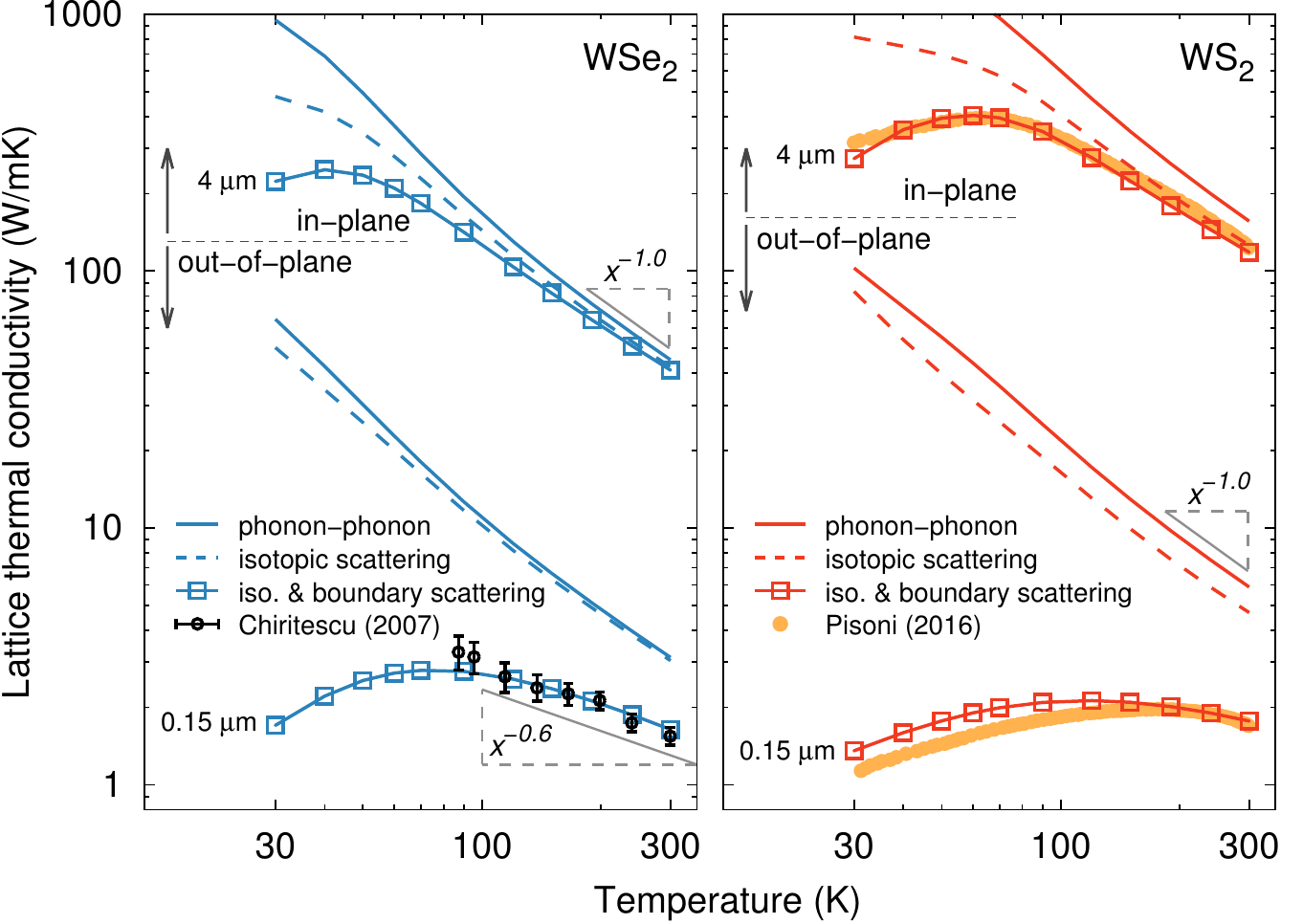}
  \caption{
    Calculated lattice thermal conductivity of WSe$_2$ and WS$_2$ in comparison with experiment.
    Solid and dashed lines and lines with squares show the calculated thermal conductivity obtained when including phonon-phonon scattering, isotopic scattering and boundary scattering.
    The intrinsic length scale $L$ is set to \unit[4]{$\mu$m} for the in-plane and \unit[0.15]{$\mu$m} for the out-of-plane conductivity.
    Experimental data for WS$_2$ and WSe$_2$ were taken from Pisoni (2016) \cite{PisJacGaa16} and Chiritescu (2007) \cite{ChiCahNgu07}, respectively.
}
  \label{fig:conductivity-WS2-WSe2}
\end{figure}

If only phonon-phonon scattering is included as a lifetime limiting mechanism in \eq{eq:matthiesen-rule}, the calculated thermal conductivity invariably exhibits a $1/T$ dependence as expected in this limit \cite{Grimvall99_chpt16} (\fig{fig:conductivity-WS2-WSe2}). Isotopic scattering lowers $\kappa$ as well as the temperature exponent in particular for temperatures below 100\,K. At room temperature the in-plane (out-of-plane) conductivity is reduced from 157 to \unit[126]{W/K\,m} (5.4 to $4.7\,\text{W/K\,m}$) in the case of WS$_2$ and from 45 to \unit[42]{W/K\,m} (3.1 to $3.0\,\text{W/K\,m}$) for WSe$_2$.

\begin{figure}
  \centering
\includegraphics[width=0.95\linewidth]{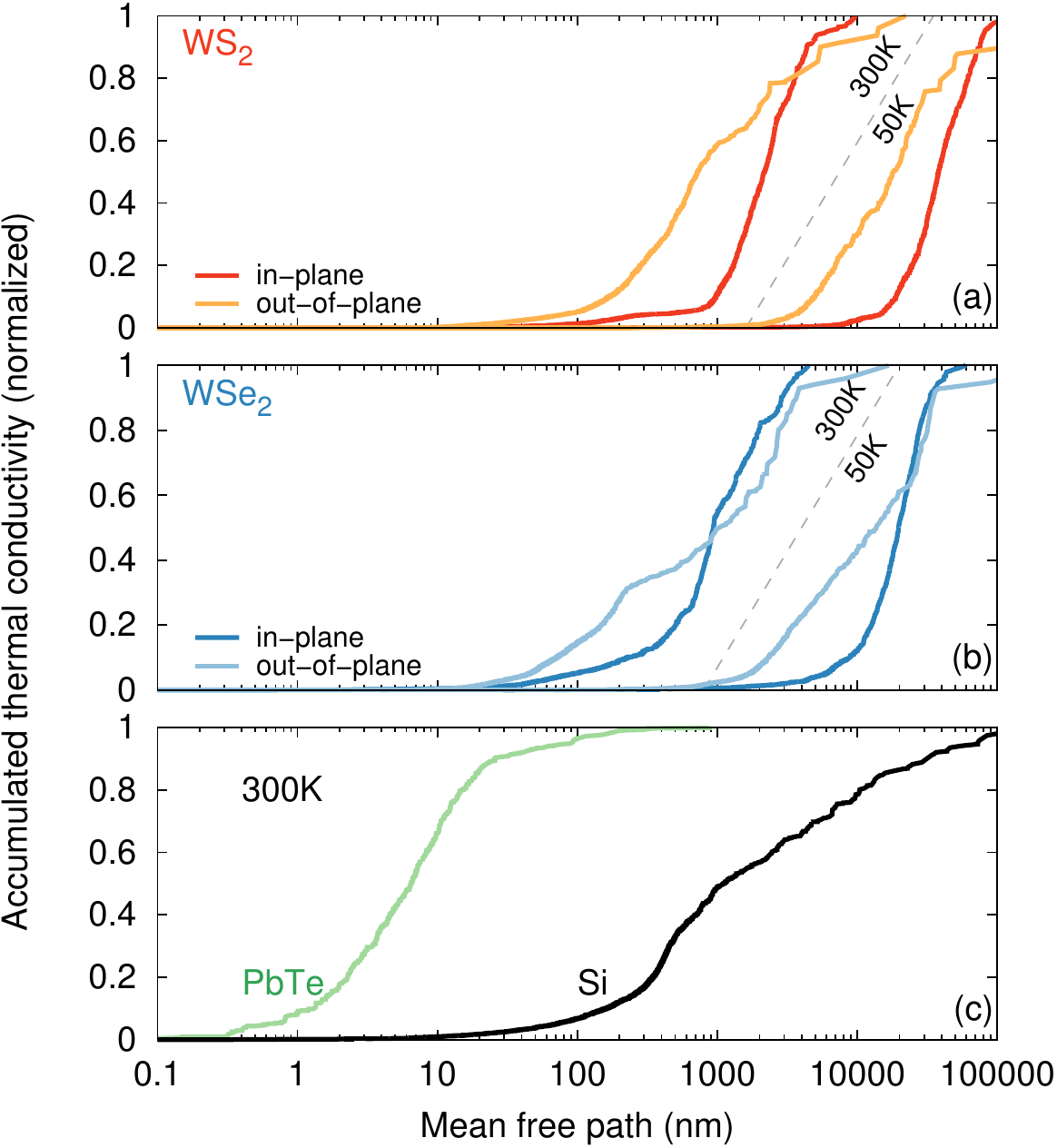}
  \caption{
    Cumulative in-plane and out-of-plane lattice thermal conductivity $\widebar{\vec{\kappa}}$ according to Eq.~\eqref{eq:cumulative-thermal-conductivity} as a function of the mode specific MFP $\Lambda_\lambda$ in (a) WS$_2$ and (b) WSe$_2$ at 50 and \unit[300]{K}, and in (c) PbTe and Si at 300\,K.
  }
  \label{fig:cumulative-conductivity}
\end{figure}

In the case of WS$_2$, the calculated in-plane conductivity at room temperature of \unit[126]{W/mk} (including phonon-phonon as well as isotopic scattering) agrees very well with the measured value of \unit[124]{W/K\,m} \cite{PisJacGaa16}. At lower temperatures there is, however, a noticeable disparity suggesting that at least one other scattering mechanism is important for $\kappa$. In fact if boundary scattering is taken into account in the form of Eq.~\eqref{eq:boundary-scattering} it is possible to reproduce the experimental in-plane conductivity over the entire temperature range using $L=4\,\mu\text{m}$.
Rather than thinking of this value as corresponding to the sample size it can be understood as a characteristic intrinsic length scale. It should also be recalled that Eq.~\eqref{eq:boundary-scattering} represents the extreme limit in which the scattering process is entirely diffusive whereas in reality some level of directional scattering can be expected \cite{CheTieWu94, NikPokAsk09}.

The notion that not only the out-of-plane \cite{ChiCahNgu07, ErhHylLin15} but also the in-plane thermal conductivity is sensitive to structural inhomogeneities is further supported by observing that the major contributions to the thermal conductivity stem from modes with MFPs of at least $1\,\mu\text{m}$ [\fig{fig:cumulative-conductivity}(a,b)], which is substantially longer than e.g., in the case of PbTe [\fig{fig:cumulative-conductivity}(c)], a system, in which nanostructuring has been used with great success to lower the thermal conductivity \cite{BisHeBlu12, TiaGarEsf12}. The representative MFP for WS$_2$ and WSe$_2$ as well as other TMDs is rather comparable to Si [\fig{fig:cumulative-conductivity}(c)], the synthesis of which ---at least currently in contrast to TMDs--- can be extremely well controlled yielding very low defect densities.

The calculated out-of-plane conductivities exhibit a considerable deviation from experiment already at room temperature (WS$_2$: $\kappa_\perp^\text{expt}=1.7\,\text{W/K\,m}$ \textit{vs.} $\kappa_\perp^\text{calc}=5.4\,\text{W/K\,m}$; WSe$_2$: $\kappa_\perp^\text{expt}=1.5\,\text{W/K\,m}$ \textit{vs.} $\kappa_\perp^\text{calc}=3.1\,\text{W/K\,m}$). Applying the same approach as in the case of the in-plane conductivity, we obtain an effective maximum MFP of $L=0.15\,\mu\text{m}$ for both materials [\fig{fig:cumulative-conductivity}(a,b)], which yields an excellent match between calculation and experiment over the entire temperature range. Of course both experiment and calculation are subject to certain errors that are difficult to control either in the form of uncertainties concerning the interpretation of the experimental raw data \cite{WilCah14} or intrinsic limitations of the theoretical description. In either case, the lower value compared to the in-plane case is consistent with the weaker binding along the $c$-axis, which implies that it is relatively easy for the material to introduce (planar) defects that reduce the effective coherence length \cite{NguBerLin10, ErhHylLin15}.

\subsection{Extension to other chalcogenides}

\begin{figure}
 \centering
\includegraphics[width=0.68\linewidth]{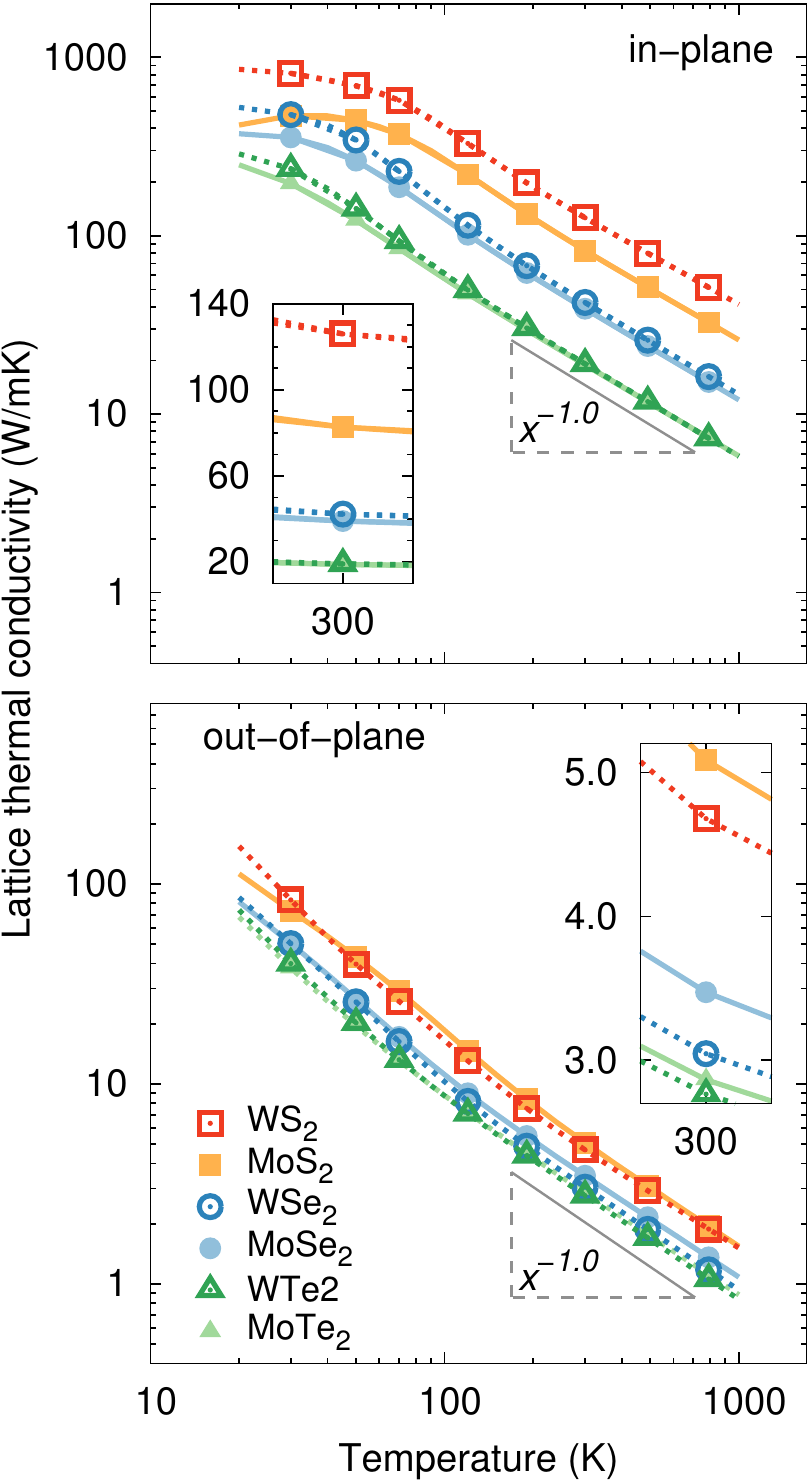}
  \caption{
    Calculated in-plane and out-of-plane lattice thermal conductivities including phonon-phonon ($\tau_{\lambda,\text{ph-ph}}$ in Eq.~\eqref{eq:matthiesen-rule}) and isotopic scattering ($\tau_{\lambda,\text{iso}}$ in Eq.~\eqref{eq:matthiesen-rule}) for the TMDs considered in this study. The insets show close-ups of the values at \unit[300]{K} on a linear scale and highlight the ordering of the materials.
  }
  \label{fig:conductivity-comparison}
\end{figure}

\begin{figure}
 \centering
\includegraphics[width=0.68\linewidth]{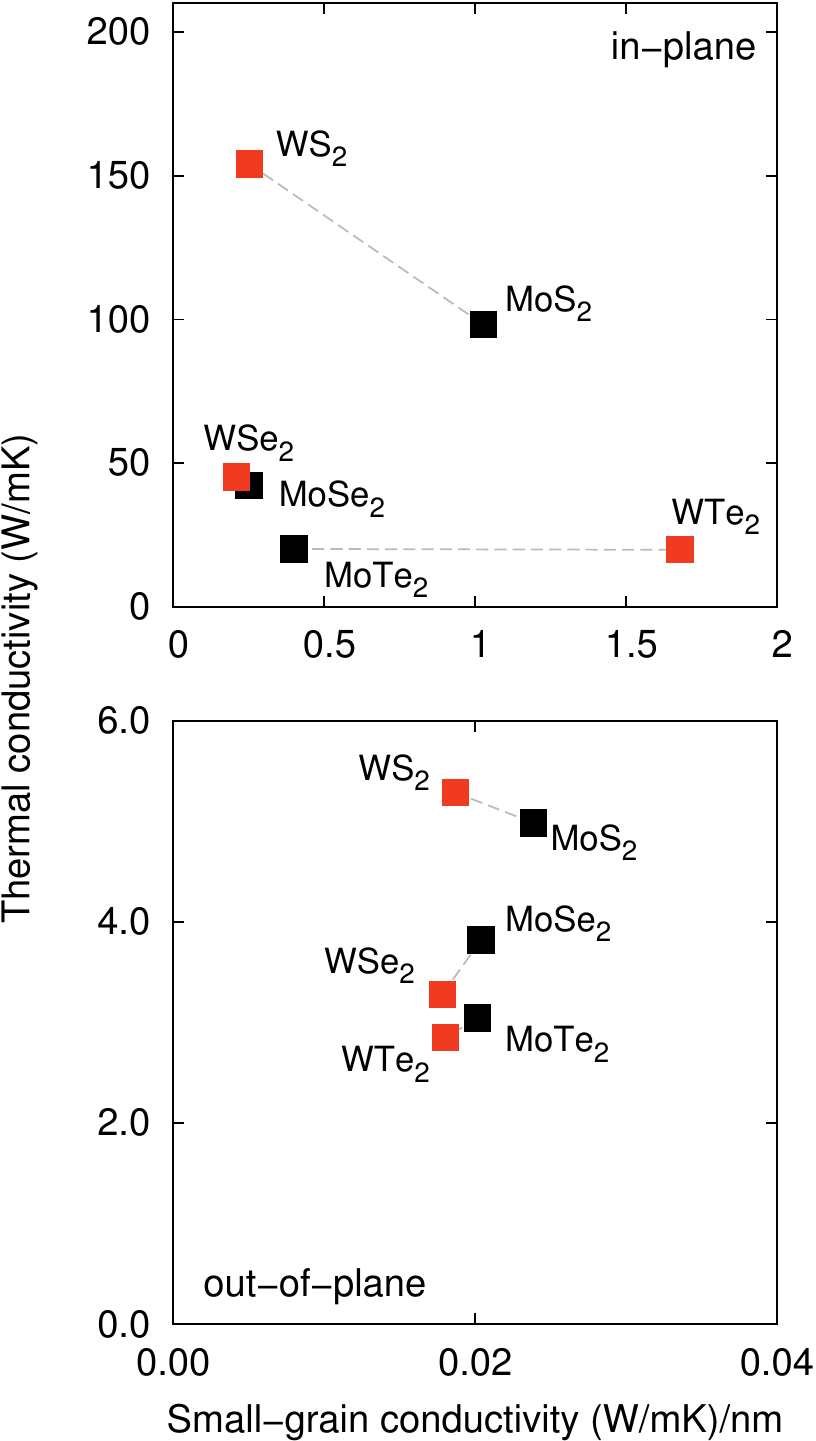}
  \caption{
    Comparison of the small-grain conductivity $\kappa_\text{sg}$ according to Eq.~\eqref{eq:small-grain-conductivity} with the respective full lattice thermal conductivity $\kappa$ according to Eq.~\eqref{eq:full-thermal-conductivity} at \unit[300]{K}.
  }
  \label{fig:small-grain-conductivity}
\end{figure}

\begin{figure}
 \centering
\includegraphics[width=0.8\columnwidth]{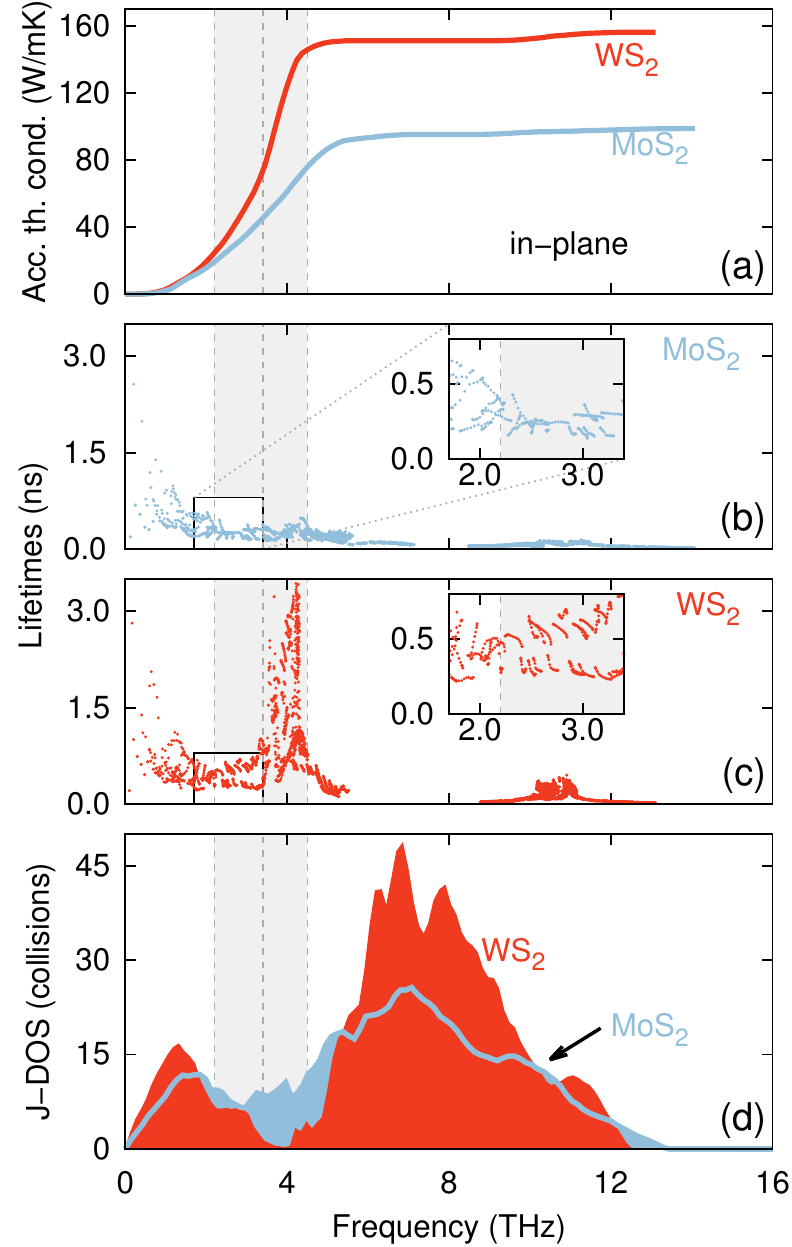}
  \caption{
    (a) Cumulative in-plane thermal conductivity as a function of the frequency of the contributing modes in MoS$_2$ and WS$_2$. 
    (b,c) Lifetimes in (b) in MoS$_2$ and (c) WS$_2$.
    (d) Weighted joint density of states for collision processes at a representative $\vec{q}$-point in the hexagonal plane.
  }
  \label{fig:comparison-MoS2-WS2}
\end{figure}

The analysis in the previous section has demonstrated both the level of accuracy of our calculations and the strong impact of impurities and other defects on many experimental measurements. These effects hinder a systematic investigation and understanding of the trends in thermal conductivity. In the following, we therefore analyze $\kappa$ for Mo and W-based TMDs considering only phonon-phonon and isotopic scattering channels.

The calculations show a systematic variation of the lattice thermal conductivity that is primarily determined by the chalcogenide species and except for the sulfides is only weakly affected by the transition metal element (\fig{fig:conductivity-comparison}). The calculated in-plane conductivities at room temperature vary from \unit[19]{W/K\,m} (MoTe$_2$, WTe$_2$) to \unit[126]{W/K\,m} (WS$_2$), while the out-of-plane data range from \unit[2.8]{W/K\,m} (WTe$_2$) to \unit[5.1]{W/K\,m} (MoS$_2$). (Recall that these values represent the limit, in which only phonon-phonon and isotopic scattering channels are available.) The thermal conductivity is thus highly anisotropic as the ratio between the in-plane and out-of-plane values ranges from 7 (MoTe$_2$) to 27 (WS$_2$) again following the sequence Te-Se-S.

The large anisotropy between in-plane and out-of-plane conductivity is largely due to the much smaller group velocities in the $c$-direction [\fig{fig:phonon-dispersions}(b)]. They are the result of the interlayer (vdW) interactions being much weaker than the intralayer (mixed covalent/ionic) bonding. This anisotropy has also been shown to give rise to a phonon focusing effect and a much lower minimum thermal conductivity than in the case of isotropic materials \cite{CheDam15}.

The chemical trend for $\kappa$ is analogous to the situation for the structural parameters, which was described in \sect{sect:description-of-vdW-solids}. The lattice parameters are the largest for the tellurides, which accordingly exhibit the smallest Brillouin zone (\fig{fig:phonon-dispersions}) and generally yield smaller group velocities resulting in lower thermal conductivities, see Eq.~\eqref{eq:full-thermal-conductivity}. One might thus be led to use the group velocities and thus the small-grain conductivity $\kappa_\text{sg}$ as a (computationally much cheaper) predictor for the thermal conductivity. A closer inspection, however, reveals no correlation between $\kappa_\text{sg}$ and the full thermal conductivity $\kappa$ (\fig{fig:small-grain-conductivity}), emphasizing the need to include phonon-phonon scattering at least at an approximate level, see e.g., Ref.~\onlinecite{BjeIveMad14}.

Compared to the other TMDs in the case of the sulfides the transition metal species has a much more pronounced effect on the in-plane lattice thermal conductivity (\fig{fig:conductivity-comparison}) with values of \unit[83]{W/K\,m} and \unit[126]{W/K\,m} for MoS$_2$ and WS$_2$, respectively. This observation is supported by experimental data as measurements for bulk MoS$_2$ fall in the range between 85 and \unit[110]{W/K\,m} \cite{LiuChoCah14} (also see \fig{fig:conductivity-MoS2}), while a value of \unit[124]{W/K\,m} was recently measured for WS$_2$ \cite{PisJacGaa16}. Since both the lattice parameters and the second-order IFCs of MoS$_2$ and WS$_2$ are similar, the differences in phonon dispersion and thus group velocities arise primarily from the mass difference between Mo and W (\fig{fig:phonon-dispersions}). The lighter mass of Mo leads to larger group velocities, which would suggest $\kappa$ to be larger for MoS$_2$, yet the opposite is the case. The difference thus must be traceable to the lifetimes.

\begin{figure}
 \centering
\includegraphics[width=0.58\linewidth]{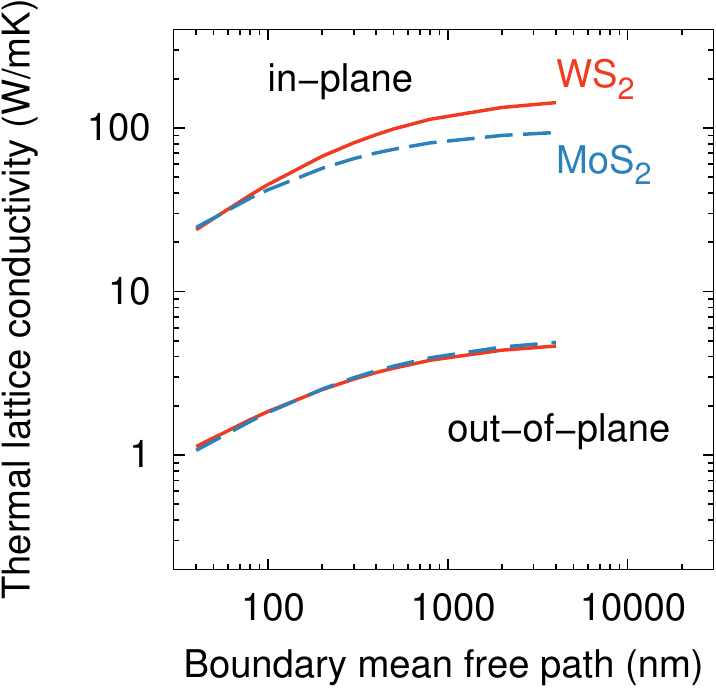}
  \caption{
    Variation of in-plane and out-of-plane thermal conductivity of MoS$_2$ and WS$_2$ with boundary scattering dimension $L$.
  }
  \label{fig:boundary-scattering-MoS2-WS2}
\end{figure}

The largest contributions to the thermal conductivity in both materials come from modes with frequencies below \unit[4.5]{THz} [\fig{fig:comparison-MoS2-WS2}(a)]. In the case of WS$_2$ the relative contributions in the interval between 2 and \unit[5]{THz} are, however, notably larger than in MoS$_2$. In fact, the lifetimes, in particular between 3.5 and \unit[4.5]{THz} are much larger in WS$_2$ [\fig{fig:comparison-MoS2-WS2}(c)] than in MoS$_2$ [\fig{fig:comparison-MoS2-WS2}(b)]. The longer lifetimes can be largely attributed to a much smaller number of allowed collision processes in this frequency range [\fig{fig:comparison-MoS2-WS2}(d)], which in turn is the direct result of the large phonon band gap in the dispersion of WS$_2$ [\fig{fig:phonon-dispersions}(a)]. The phonon gaps in these materials are caused by the mass difference between cation and anion species, which is the largest for WS$_2$ among the TMDs considered in this work. If boundary scattering is included the relative importance of phonon-phonon scattering is reduced, which diminishes the difference between WS$_2$ and MoS$_2$ (\fig{fig:boundary-scattering-MoS2-WS2}).

Differences in lifetimes between Mo and W-based TMDs, albeit smaller than for the sulfides, are also present for the selenides and tellurides. In these materials the lifetime effect is, however, outweighed by the group velocity contribution (compare the insets in \fig{fig:conductivity-comparison}). The present analysis nonetheless demonstrates the importance of lifetime effects for understanding the thermal conductivity in these materials.

\section{Conclusions}

In the present work we investigated finite temperature properties as well as the lattice thermal conductivity in Mo and W-based TMDs employing a combination of density functional and Boltzmann transport theory. The calculations were carried out using the vdW-DF-CX functional, which was shown to yield excellent agreement with experimental lattice constants at room temperature with an average relative error below 0.2\%\ (\tab{tab:structural-parameters-TMDs}).

The calculated in-plane conductivities at room temperature are in good agreement with experimental data for high-purity material, when only phonon-phonon and isotopic scattering are included (Figs.~\ref{fig:conductivity-MoS2} and \ref{fig:conductivity-WS2-WSe2}). Explaining the experimental data over the entire temperature, however, requires inclusion of at least one additional scattering mechanism (here boundary scattering) that limits the phonon MFP (\fig{fig:conductivity-WS2-WSe2}). The latter effect is even more pronounced in the case of the out-of-plane conductivity, for which we obtain an intrinsic length scale of $L=0.15\,\mu\text{m}$ to be compared with $L=4\mu\text{m}$ in the in-plane situation.

The sensitivity of the thermal conductivity to structural inhomogeneities can be explained in terms of the long MFP of the modes that contribute the most strongly to $\kappa$ (\fig{fig:cumulative-conductivity}). The MFP of these modes (including phonon-phonon and isotopic scattering) is at least $1\,\mu\text{m}$, which is comparable to silicon but much larger than e.g., PbTe. This behavior is promising for thermoelectric applications, where lowering the lattice part of the thermal conductivity is a widely employed approach for increasing the thermodynamic efficiency. On the other hand, it can pose problems for electronic and optoelectronic applications, which require a large $\kappa$ for rapid heat dissipation.

A comprehensive analysis shows that the thermal conductivity is primarily affected by the chalcogenide species and increases in the order Te-Se-S (\fig{fig:conductivity-comparison}). As expected from the elemental masses, MoTe$_2$ and MoSe$_2$ exhibit a higher conductivity than the respective W-based TMDs. For the sulfides the situation is inverted, which can be traced to the larger phononic band gap in the case of WS$_2$ (see Figs.~\ref{fig:phonon-dispersions} and \ref{fig:comparison-MoS2-WS2}). This observation suggests that in principle phonon-engineering can be achieved not only via the group velocity term in Eq.~\eqref{eq:full-thermal-conductivity} and microstructuring but also via the phonon-phonon scattering.

The present study provides a comprehensive set of lattice thermal conductivities for bulk TMDs that establishes bounds set by phonon-phonon scattering and intrinsic length scales. It thereby forms the basis for future studies on these systems, which could focus e.g., on vdW solids comprising different layers.

\begin{acknowledgments}
We gratefully acknowledge fruitful discussions with Per Hyldgaard. This work has been supported by the Knut and Alice Wallenberg foundation and through computer time allocations by the Swedish National Infrastructure for Computing at NSC (Link\"oping) and PDC (Stockholm).
\end{acknowledgments}

\end{document}